\documentclass[english]{article}
\usepackage[T1]{fontenc}
\usepackage[latin9]{inputenc}
\usepackage{verbatim}
\usepackage{float}
\usepackage{amsmath}
\usepackage{graphicx}
\usepackage{esint}
\usepackage{amsfonts}
\usepackage{hyperref}

\usepackage{amssymb}
\usepackage{latexsym}
\usepackage{epsfig}
\usepackage{enumerate}

\makeatletter

\newcommand{\be}{\begin{equation}}
\newcommand{\ee}{\end{equation}}

\allowdisplaybreaks

\makeatother

\usepackage{babel}
\begin{document}
{}~ \hfill\vbox{\hbox{CTP-SCU/2019008}}\break
\vskip 3.0cm
\centerline{\Large \bf  Are nonperturbative AdS vacua possible in bosonic string theory?}

\vspace*{10.0ex}
\centerline{\large Peng Wang, Houwen Wu and Haitang Yang}
\vspace*{7.0ex}
\vspace*{4.0ex}
\centerline{\large \it College of physics}
\centerline{\large \it Sichuan University}
\centerline{\large \it Chengdu, 610065, China} \vspace*{1.0ex}
\vspace*{4.0ex}

\centerline{pengw@scu.edu.cn, iverwu@scu.edu.cn, hyanga@scu.edu.cn}
\vspace*{10.0ex}
\centerline{\bf Abstract} \bigskip \smallskip
In this paper, following the work of Hohm and Zwiebach [arXiv:1905.06583], we show that in bosonic string theory nonperturbative anti-de Sitter (AdS) vacua could exist with all $\alpha^{\prime}$ corrections included. We also discuss the possibility of the coexistence of  nonperturbative dS and AdS vacua.

\vfill
\eject
\baselineskip=16pt
\vspace*{10.0ex}

Whether bosonic string theory permits stable de Sitter (dS) or anti-de Sitter (AdS) vacua is a long-standing unsolved problem.
There are conjectures that superstring theory does not have solutions of dS vacua \cite{Obied:2018sgi,Agrawal:2018own,Garg:2018reu}.
Another conjecture states that there is no stable nonsupersymmetric AdS vacuum with fluxes \cite{Ooguri:2016pdq}. Incredibly, by analyzing the nonperturbative properties of the spacetime action of closed string theory (also known as the low-energy effective string theory), Hohm and Zwiebach \cite{Hohm:2019ccp,Hohm:2019jgu}
recently showed that \emph{nonperturbative} dS vacua are possible in bosonic string theory. The most important ingredients in their arguments are   $O(d,d)$ symmetry  and the classification of all of the $\alpha'$ corrections for particular configurations.

It is well known from the work of  Meissner and Veneziano  \cite{Veneziano:1991ek} in 1991 that, at the zeroth order of $\alpha'$, when all  fields depend only on time, the $D=d+1$-dimensional spacetime action of closed string theory reduces to an $O(d,d)$-invariant \emph{reduced action}. Soon after that, in Refs. \cite{Sen:1991zi,Sen:1991cn} Sen extended this result  to full string field theory. Specifically, by considering an exact solution of the string field that was independent of $m$-dimensional spacetime coordinates ($m\le d$)\footnote{We concentrate on noncompact configurations here.} , Sen proved the following. (i) The space of such solutions has an $O(m,m)$ symmetry. In the language of low energy effective theory, the reduced action derived from such solutions   possesses an $O(m,m)$ symmetry to all orders in $\alpha'$. (ii) The $m$ coordinates could be all spacelike or include one timelike coordinate, as explained  in Ref. \cite{Sen:1991cn}. (iii) In the solution space, inequivalent solutions are connected by nondiagonal $O(m)\otimes O(m)$ transformations [$O(m-1,1)\otimes O(m-1,1)$ if one of the $m$ coordinates is timelike]. (iv) Other generators of $O(m,m)$ outside of the nondiagonal  $O(m)\otimes O(m)$ [or $O(m-1,1)\otimes O(m-1,1)$] generate gauge transformations accompanied by a shift of the dilaton, and thus equivalent solutions. On the other hand, in Ref. \cite{Meissner:1991zj}, from the perspective of  $\sigma$ model expansion, since the nilpotency of the BRST operator $Q$ is not altered by an $O(d,d)$ transformation, it was argued that the  $O(d,d)$ symmetry should persist at all orders in $\alpha'$ for the reduced action. It is expected that, in terms of the standard fields, the $O(d,d)$ transformations  receive higher-order $\alpha'$ corrections when introducing higher-derivative terms to the reduced action. For configurations depending only on time, to the first order in $\alpha'$, in Ref. \cite{Meissner:1996sa}, Meissner demonstrated that one can trade it with standard $O(d,d)$ transformations in terms of $\alpha'$-corrected fields. In the appendix we show that this is also true for configurations that only depend on one spatial coordinate $x$, i.e., the case we study in this paper.

As for the yet unknown higher-order $\alpha'$ corrections, some important progress has been made recently using the formalism of double field theory \cite{Hohm:2013jaa,Hohm:2014xsa,Marques:2015vua,Hohm:2016lge,Hohm:2015doa,Baron:2018lve}. Remarkably, in Ref. \cite{Hohm:2019jgu, Hohm:2015doa} Hohm and Zwiebach demonstrated that, for cosmological, purely time-dependent configurations, the $O(d,d)$-covariant closed string spacetime action can be expressed in a very simple form. All orders of  $\alpha^{\prime}$ corrections do not include
the trivial dilaton and can be constructed using even powers of
$\partial_{t}\mathcal{S}$, where $\mathcal{S}$ is the spatial part
of the generalized metric defined in Eq. (\ref{M}). This surprising simplification of the $\alpha^{\prime}$ corrections enabled them to discuss the nonperturbative solutions.  The most interesting result they obtained is that nonperturbative dS vacua are possible for bosonic string theory \cite{Hohm:2019ccp,Hohm:2019jgu}, which possibly provides a cornerstone for the connection between string theory and our real world. In this paper, following their derivations, we show that nonperturbative AdS vacua are also possible with all  $\alpha'$ corrections for bosonic string theory.

It is worth noting that we work in the string frame and not the Einstein frame, the same in Hohm and Zwiebach's work \cite{Hohm:2019ccp,Hohm:2019jgu}. It is still unclear if there could be dS or AdS solutions in the Einstein frame, since when we substitute the
solutions with a constant dilaton field $\phi$ and Hubble parameter
$\bar{H}_{0}$ back into the Einstein frame, $\bar{H}_{0}^{E}$ goes to
zero and the metric becomes flat.    Another issue is that in order to completely determine the dS/AdS vacua, we still need to know all of the $\alpha'$ corrections. One of the purposes of this paper is to deny the nonexistence of AdS vacua in bosonic string theory, rather than provide exact solutions.


For the sake of completeness, let us briefly summarize Hohm and Zwiebach's work on  nonperturbative dS vacua. Details can be found in Ref. \cite{Hohm:2019jgu}. To the zeroth order of $\alpha'$, the $D=d+1$-dimensional spacetime action of closed string theory is

\begin{equation}
I_{0}\equiv\int d^{D}x\sqrt{-g}e^{-2\phi}\left[R+4\left(\partial_{\mu}\phi\right)^{2}-\frac{1}{12}H_{ijk}H^{ijk}\right],
\end{equation}

\noindent where $g_{\mu\nu}$ is the string metric, $\phi$ is the
dilaton and $H_{ijk}=3\partial_{\left[i\right.}b_{\left.jk\right]}$
is the field strength of the antisymmetric Kalb-Ramond $b_{ij}$
field. For cosmological backgrounds, choosing the synchronous
gauge $g_{tt}=-1$, $g_{ti}=b_{t\mu}=0$,

\begin{equation}
g_{\mu\nu}=\left(\begin{array}{cc}
-1 & 0\\
0 & G_{ij}\left(t\right)
\end{array}\right),\qquad b_{\mu\nu}=\left(\begin{array}{cc}
0 & 0\\
0 & B_{ij}\left(t\right)
\end{array}\right),\qquad\phi=\phi\left(t\right),
\end{equation}

\noindent and defining the $O\left(d,d\right)$ dilaton $\Phi$ as

\noindent
\begin{equation}
e^{-\Phi}=\sqrt{g}e^{-2\phi},
\end{equation}

\noindent the action can be rewritten as

\noindent
\begin{equation}
I_{0}=\int dte^{-\Phi}\left[-\dot{\Phi}^{2}-\frac{1}{8}\mathrm{Tr}\left(\dot{\mathcal{S}}^{2}\right)\right],\label{ST action}
\end{equation}

\noindent with

\noindent
\begin{equation}
M=\left(\begin{array}{cc}
G^{-1} & -G^{-1}B\\
BG^{-1} & G-BG^{-1}B
\end{array}\right),\qquad\mathcal{S}=\eta M=\left(\begin{array}{cc}
BG^{-1} & G-BG^{-1}B\\
G^{-1} & -G^{-1}B
\end{array}\right),\label{M}
\end{equation}

\noindent where $M$, a $2d\times 2d$ matrix, is the spatial part of the generalized metric $\mathcal{H}$ of
double field theory,   $\dot{A}\equiv\partial_{t}A$, and $\eta$ is the invariant metric of the $O\left(d,d\right)$
group

\noindent
\begin{equation}
\eta=\left(\begin{array}{cc}
0 & I\\
I & 0
\end{array}\right).
\end{equation}

\noindent Noticing that $M$ is symmetric and $\mathcal{S}=\mathcal{S}^{-1}$,  this action
is manifestly invariant under the $O\left(d,d\right)$ transformations

\noindent
\begin{equation}
\Phi\longrightarrow\Phi,\qquad\mathcal{S}\longrightarrow\tilde{\mathcal{S}}=\Omega^{T}\mathcal{S}\Omega,\label{O(d,d) trans}
\end{equation}

\noindent where $\Omega$ is a constant matrix, satisfying

\noindent
\begin{equation}
\Omega^{T}\eta\Omega=\eta.
\end{equation}

\noindent If we choose the Friedmann-Lemaitre-Robertson-Walker (FLRW) metric, $G_{ij}=\delta_{ij}a^{2}\left(t\right)$, and a
vanishing Kalb-Ramond field $B=0$:

\begin{equation}
ds^{2}=-dt^{2}+a^{2}\left(t\right)\delta_{ij}dx^{i}dx^{j},
\label{FLRW}
\end{equation}

\noindent the matrix $\mathcal{S}$ becomes

\noindent
\begin{equation}
\mathcal{S}=\left(\begin{array}{cc}
0 & G\\
G^{-1} & 0
\end{array}\right).
\end{equation}

\noindent Applying the $O(d)\otimes O(d)$ transformation, we
obtain a new solution,

\begin{equation}
\tilde{\mathcal{S}}=\left(\begin{array}{cc}
0 & G^{-1}\\
G & 0
\end{array}\right),
\end{equation}

\noindent which implies that the action is invariant under $a\left(t\right)\rightarrow a^{-1}\left(t\right)$, which is known as the scale-factor duality in traditional string
cosmology. The next step is to include the $\alpha^{\prime}$
corrections. Benefiting from the $O\left(d,d\right)$ invariance, the corrections
are classified by  even powers of $\dot{\mathcal{S}}$ only:

\begin{eqnarray}
I & = & \int d^{D}x\sqrt{-g}e^{-2\phi}\left(R+4\left(\partial\phi\right)^{2}-\frac{1}{12}H^{2}+\frac{1}{4}\alpha^{\prime}\left(R^{\mu\nu\rho\sigma}R_{\mu\nu\rho\sigma}+\ldots\right)+{\alpha'}^2(\ldots)+\ldots \right),\label{eq:original action with alpha}\\
 & = & \int dte^{-\Phi}\left(-\dot{\Phi}^{2}+ {\sum_{k=1}^{\infty}}\left(\alpha^{\prime}\right)^{k-1}c_{k}\mathrm{tr}\left(\dot{\mathcal{S}}^{2k}\right)\right).\label{eq:Odd action with alpha}
\end{eqnarray}

\noindent Eq. (\ref{eq:original action with alpha}) is the action for the general background
with all $\alpha^{\prime}$ corrections. Eqn (\ref{eq:Odd action with alpha})  is
the $O\left(d,d\right)$-covariant action  applied to the metric (\ref{FLRW}) with  $B=0$,
where $c_{1}=-\frac{1}{8}$
to recover Eq. (\ref{ST action}) and $c_{k\geq2}$ are
undetermined constants. Using the action   (\ref{eq:Odd action with alpha}), in Refs. \cite{Hohm:2019ccp,Hohm:2019jgu} Hohm and Zwiebach showed that nonperturbative
dS vacua are permitted for infinitely many classes of  $c_{k\geq2}$, namely, $a(t)=e^{H_0 t}$ with $H_0\not= 0$.

Now we want to investigate if nonperturbative AdS vacua are also allowed. To address this question, an appropriate ansatz is crucial. We take the ansatz

\begin{equation}
ds^{2}=-a^{2}\left(x\right)dt^{2}+dx^{2}+a^{2}\left(x\right)\left(dy^{2}+dz^{2}+\ldots\right),\label{eq:BH metric}
\end{equation}

\noindent whose metric components depend on a single space direction, say, $x$. The dimensionality is still $D=d+1$. As we explained earlier, in Refs. \cite{Sen:1991zi,Sen:1991cn}  Sen proved that the reduced action based on such solutions also maintains $O(d,d)$ symmetry to all orders in $\alpha'$. The coset for the space of such solutions is $O(d-1,1)\otimes O(d-1,1)/O(d-1,1)$, in contrast to $O(d)\otimes O(d)/O(d)$ for the cosmological solutions (\ref{FLRW}). In the Appendix we explicitly show that for this ansatz the spacetime action (\ref{eq:original action with alpha}) also possesses the standard $O(d,d,R)$ symmetry and can be reduced to

\begin{equation}
\bar{I}= -\int dxe^{-\Phi}\left(-\Phi^{\prime2}+ {\sum_{k=1}^\infty}\left(\alpha^{\prime}\right)^{k-1}\bar{c}_{k}\mathrm{tr}\left(\mathcal{M}^{\prime2k}\right)\right),\label{eq: Odd with alpha x}
\end{equation}

\noindent where $A^{\prime}\equiv\partial_{x}A$. Note the overall minus sign and the fact that we have a new set of undetermined coefficients $\bar c_k$ other than the $c_k$'s in Eq. (\ref{eq:Odd action with alpha}).  It turns out that

\begin{equation}
\bar{c}_{2k-1}=c_{2k-1},\quad\quad \bar{c}_{2k}=-c_{2k},\quad\quad \mathrm{for}\quad k=1,2,3\ldots \label{eq:coeff relation}
\end{equation}

\noindent and

\begin{equation}
\mathcal{M}=\left(\begin{array}{cccc}
0 & 0 & -a^{2}\left(x\right) & 0\\
0 & 0 & 0 & a^{2}\left(x\right)\delta_{ij}\\
-a^{-2}\left(x\right) & 0 & 0 & 0\\
0 & a^{-2}\left(x\right)\delta_{ij} & 0 & 0
\end{array}\right),
\end{equation}

\noindent where $i,j=y,z,\ldots$. The equations of motion (EOM) of (\ref{eq: Odd with alpha x}) can be calculated directly,

\begin{eqnarray}
\Phi^{\prime\prime}+\frac{1}{2}\bar{H}\bar{f}\left(\bar{H}\right) & = & 0,\nonumber \\
\frac{d}{dx}\left(e^{-\Phi}\bar{f}\left(\bar{H}\right)\right) & = & 0,\nonumber \\
(\Phi^{\prime})^2+\bar{g}\left(\bar{H}\right) & = & 0,\label{eq:EOM}
\end{eqnarray}

\noindent where

\begin{eqnarray}
\bar{H}\left(x\right) & = & \frac{a^{\prime} \left(x\right)}{a\left(x\right)},\nonumber \\
\bar{f}\left(\bar{H}\right) & = & d{\sum_{k=1}^\infty}\left(-\alpha^{\prime}\right)^{k-1} 2^{2\left(k+1\right)}k\bar{c}_{k}\bar{H}^{2k-1},\nonumber \\
\bar{g}\left(\bar{H}\right) & = & d {\sum_{k=1}^\infty} \left(-\alpha^{\prime}\right)^{k-1} 2^{2k+1}\left(2k-1\right)\bar{c}_{k} \bar{H}^{2k}.\label{eq:EOM fh gh}
\end{eqnarray}

\noindent It is easy to see that $\bar g'(\bar H)=\bar H \bar f'(\bar H)$. Note that $\bar{H}\left(x\right)$ is not the Hubble parameter since our background is space dependent.

Now, let us check whether there is a solution $a^{2}\left(x\right)=e^{2\bar{H}_{0}x}$
for the EOM (\ref{eq:EOM}) such that

\begin{equation}
ds^{2}=-e^{2\bar{H}_{0}x}dt^{2}+dx^{2}+e^{2\bar{H}_{0}x}\left(dy^{2}+dz^{2}+\ldots\right).\label{eq:ads solution}
\end{equation}

\noindent The scalar curvature of this metric is

\begin{equation}
R=-D\left(D-1\right)\bar{H}_{0}^{2},
\label{eq:Ricci Scalar}
\end{equation}

\noindent which implies that the metric (\ref{eq:ads solution}) is
an AdS background for constant  $\bar{H}_{0}\neq0$. To see this more clearly, we apply the transformation $x \to - \log[\bar H_0 \xi]/\bar H_0$ and recover the familiar Poincare coordinate

\begin{equation}
ds^2 = \frac{1/\bar H_0^2}{\xi^2}\big(-dt^2 +d\xi^2 + dy^2+dz^2+\ldots\big).
\end{equation}
So, we have  $\bar H_0=1/R_{AdS}$, consistent with Eq.(\ref{eq:Ricci Scalar}). If we do not include the $\alpha^{\prime}$ corrections,
$\bar{f}\left(\bar{H}\right)\sim \bar H_0$ is a constant. From the second equation
of (\ref{eq:EOM}), we can figure out that $\Phi$ is a constant.
Therefore, to satisfy the first equation of (\ref{eq:EOM}), we must have
$\bar{H}_{0}=0$, and thus the metric (\ref{eq:ads solution}) becomes
flat and there is no AdS solution. One thus concludes that there
is no $D$-dimensional AdS vacuum without fluxes and $\alpha^{\prime}$
corrections.

Our aim is to search for solutions with constant $\bar{H}_{0}\neq0$.  Considering
the effects of $\alpha^{\prime}$ corrections, if there is a nonvanishing
$\bar{H}_{0}$ solution, $\bar{f}\left(\bar{H}_{0}\right)$ is
a constant and then $\Phi$ is also a constant from the second equation
of (\ref{eq:EOM}). Finally, from the first and third equations of (\ref{eq:EOM}), we  obtain the condition for a nonvanishing $\bar{H}_{0}$ solution:

\begin{equation}
\bar{f}\left(\bar{H}_{0}\right)=\bar{g}\left(\bar{H}_{0}\right)=0.\label{eq:fh condition}
\end{equation}

Let us determine the general form of $\bar{f}\left(\bar{H}\right)$ with specific choices for $c_{k\geq2}$
that satisfy the condition (\ref{eq:fh condition}). Instead of  $\bar{f}\left(\bar{H}\right)$,
it is better to consider its integral,

\begin{equation}
\bar{F}\left(\bar{H}\right)\equiv\int_{0}^{\bar{H}}f\left(\bar{H}^{\prime}\right)d\bar{H}^{\prime}.
\end{equation}

\noindent The condition $\bar{f}\left(\bar{H}_{0}\right)=\bar{g}\left(\bar{H}_{0}\right)=0$
is replaced by

\begin{equation}
\bar{F}\left(\bar{H}_{0}\right)=\bar{F}^{\prime}\left(\bar{H}_{0}\right)=0.
\end{equation}

\noindent It is then easy to understand that the general form

\begin{equation}
\bar{F}\left(\bar{H}\right)=-d\bar{H}^{2}\left(1+ {\sum_{p=1}^{\infty}}\bar{d}_{p}\left(\alpha^{\prime}\right)^{p} \bar{H}^{2p}\right){\prod_{i=1}^k}\left(1-\left(\frac{\bar{H}}{\bar{H}_{0}^{\left(i\right)}}\right)^{2}\right)^{2}, \label{eq:Fform}
\end{equation}

\noindent admits  $2k$ solutions of AdS vacua: $\pm\bar{H}_{0}^{\left(1\right)},\ldots,\pm\bar{H}_{0}^{\left(k\right)}$, for an arbitrary integer $k>0$. So, the question is: do the coefficients $\bar c_k$ support the functional form (\ref{eq:Fform}) ?  Although it appears impossible to obtain a definite answer, the bottom line is that  $\alpha^{\prime}$ corrections do support the possibility of nonperturbative AdS vacua.  It is worth noting that ``nonperturbative''
here means that we  use all $\alpha^{\prime}$ corrections to
obtain the solution but not to obtain the solution from the two-derivative
equations and then be $\alpha^{\prime}$ corrected.
There  exists another ``stronger'' version of ``nonperturbative'', namely
$\bar{F}\left(\bar{H}\right)$ cannot be expressed by a series expansion of $\alpha'$  \cite{Krishnan:2019mkv}. The same scenario occurs for the dS vacua as explained in Refs. \cite{Hohm:2019ccp,Hohm:2019jgu}.


However, the real story may   be intriguing. As an illustration, let us assume that all orders of $\alpha^{\prime}$ corrections  have a very special form that gives

\begin{equation}
\bar{f}\left(\bar{H}\right)=-\frac{2d}{\sqrt{\alpha^{\prime}}}\sin\left(\sqrt{\alpha^{\prime}}\bar{H}\right)=-2d {\sum_{k=1}^{\infty}}\left(\alpha^{\prime}\right)^{k-1}\frac{1} {\left(2k-1\right)!}H^{2k-1}.\label{eq:fh sin}
\end{equation}

\noindent This functional form is a valid candidate for Eq. (\ref{eq:EOM fh gh}) for
special choices of $\bar c_{k\geq2}$. It includes all orders of $\alpha'$ corrections and evidently is nonperturbative.
The solutions  satisfying the condition $\bar{f}\left(\bar{H}_{0}\right)=0$ are
\begin{equation}
\sqrt{\alpha^{\prime}}\bar{H}_{0}=2\pi,4\pi,\ldots
\end{equation}

\noindent It is easy to check that $\bar{g}\left(\bar{H}_{0}\right)=0$ is also satisfied for these solutions, leading to a discrete infinity of AdS vacua. However, note the coefficient relations (\ref{eq:coeff relation}) between dS from Eq. (\ref{eq:Odd action with alpha}) and AdS from Eq. (\ref{eq: Odd with alpha x}):
$\bar{c}_{2k+1}=c_{2k+1}$, $\bar{c}_{2k}=-c_{2k}$. We can immediately see that if $\bar{f}\left(\bar{H}\right)\sim\sin\left(\sqrt{\alpha^{\prime}}\bar{H}\right)$ for AdS, then the corresponding function
$f\left(H\right)\sim\sinh\left(\sqrt{\alpha^{\prime}}H\right)$ in dS, and vice versa. But the $sinh$ function has no nontrivial zero. So,  for the trial function (\ref{eq:fh sin}), AdS or dS vacua cannot coexist and only one of them survives.

This looks like merely a coincidence since
in any case, one could use a general form of Eq. (\ref{eq:Fform}) to
permit  AdS or dS vacua. But we have some reasons to conjecture that by plugging the dS (AdS) metric into the yet unknown
infinite $\alpha^{\prime}$ expansion, one could sum the series
into an expression including a  factor that is very close to the trial function of Eq. (\ref{eq:fh sin})
\footnote{We want to emphasize that the real functional form of $\bar{f}\left(\bar{H}\right)$ could be more complicated than Eq. (\ref{eq:fh sin}). We simply use this toy model to discuss the coexistence of nonperturbative dS and AdS vacua.}. In Ref. \cite{Wang:2017mpm}, we showed that,
when expressed in Riemann normal coordinates, the AdS (dS) metric
can be expressed in a simple form, which is called the $J$-factor by some mathematicians.
To see this explicitly, by considering the nonlinear sigma model of
string theory

\begin{equation}
S=-\frac{1}{4\pi\alpha'}\int_{\Sigma}g_{ij}(X)\partial_{\alpha}X^{i}\partial^{\alpha}X^{j},
\end{equation}

\noindent we can expand $X^{i}$ at some point $\bar{x}$, say, $X^{i}\left(\tau,\sigma\right)=\bar{x}^{i}+\sqrt{\alpha^{\prime}}\mathbb{Y}^{i}\left(\tau,\sigma\right)$,
where the $\mathbb{Y}^{i}$'s are dimensionless fluctuations. Locally
around any point, one can always pick Riemann normal coordinates

\begin{eqnarray}
g_{ij}\left(X\right) & = & \eta_{ij}+\frac{\ell_{s}^{2}}{3}R_{iklj}\mathbb{Y}^{k}\mathbb{Y}^{l}+\frac{\ell_{s}^{3}}{6}D_{k}R_{ilmj}\mathbb{Y}^{k}\mathbb{Y}^{l}\mathbb{Y}^{m}\nonumber \\
 &  & +\frac{\ell_{s}^{4}}{20}\left(D_{k}D_{l}R_{imnj}+\frac{8}{9}R_{iklp}R_{\;mnj}^{p}\right)\mathbb{Y}^{k}\mathbb{Y}^{l}\mathbb{Y}^{m}\mathbb{Y}^{n}+\ldots.
\end{eqnarray}

\noindent When the background is  maximally symmetric, the expansion is greatly simplified and can
be summed  into a closed form.  For dS, we have

\begin{equation}
S_{dS}=-\frac{1}{4\pi}\int_{\Sigma}\partial\mathbb{Y}^{i}\partial\mathbb{Y}^{j}\left[\frac{\sin^{2}\left(\frac{\sqrt{\alpha'}}{R_{dS}}\mathbb{W}\right)}{\left(\frac{\sqrt{\alpha'}}{R_{dS}}\mathbb{W}\right)^{2}}\right]^{a}\,_{i}\,\eta_{aj}\,,\qquad\left(\mathbb{W}^{2}\right)_{\quad b}^{a}\equiv\delta_{b}^{a}\mathbb{Y}^{2}-\mathbb{Y}^{a}\mathbb{Y}_{b}.
\end{equation}

\noindent If the background is AdS, we get

\begin{equation}
S_{AdS}=-\frac{1}{4\pi}\int_{\Sigma}\partial\mathbb{Y}^{i}\partial\mathbb{Y}^{j}\left[\frac{\sinh^{2}\left(\frac{\sqrt{\alpha'}}{R_{AdS}}\mathbb{W}\right)}{\left(\frac{\sqrt{\alpha'}}{R_{AdS}}\mathbb{W}\right)^{2}}\right]^{a}\,_{i}\,\eta_{aj}\,,\qquad\left(\mathbb{W}^{2}\right)_{\quad b}^{a}\equiv\delta_{b}^{a}\mathbb{Y}^{2}-\mathbb{Y}^{a}\mathbb{Y}_{b}.
\end{equation}

\noindent Noting that $H_{0}\sim1/R_{dS}$ and $\bar{H}_{0}\sim1/R_{AdS}$,
the results strongly suggest that the beta functions or EOMs of
these two actions $S_{dS}$ and $S_{AdS}$ may behave very similarly to $f\left(H\right)\sim\sin\left(\sqrt{\alpha^{\prime}}H\right)$
and $\bar{f}\left(\bar{H}\right)\sim\sinh\left(\sqrt{\alpha^{\prime}}\bar{H}\right)$,
or, equivalently speaking, there are  nonperturbative dS vacua but not
nonperturbative AdS vacua, or vice versa. So it looks like we still need more information about the $\alpha'$ corrections to give a definite answer.

Finally, we wish to remark that we have only considered the string
metric. The relation between the Einstein metric $g_{\mu\nu}^{E}$ and
string metric $g_{\mu\nu}$ is $g_{\mu\nu}^{E}=e^{-\frac{4\phi}{D-2}}g_{\mu\nu}$.
When we substitute our solution with a constant $\phi$ and $\bar{H}_{0}= 0$
back into the Einstein frame, $\bar{H}_{0}^{E}$  goes to zero and the metric
becomes flat. This implies that there is no dS or AdS vacuum when $\phi$
is a constant in the Einstein frame without $\alpha'$ corrections.

\vspace{5mm}

\noindent {\bf Acknowledgements}
We are deeply indebted to Olaf Hohm and Barton Zwiebach for illuminating discussions and advice.  We are also  grateful to Hiroaki Nakajima, Bo Ning, Shuxuan Ying for very helpful discussions and suggestions. This work is supported in part by the National Natural Science Foundation of China (Grants No. 11875196, 11375121 and 11005016). 

\section*{Appendix}

This appendix has two purposes. The first is to explicitly show that, at the leading order in $\alpha'$, for our ansatz (\ref{eq:BH metric}) the $O(d,d)$ symmetry of the spacetime action can be expressed in the standard form in terms of $\alpha'$-corrected fields. The derivations follow the same pattern  as the calculations in Refs. \cite{Veneziano:1991ek,Meissner:1996sa}, except for some minus signs in particular places that account for the difference between time and space coordinates.

The second purpose is to briefly demonstrate that, based on our ansatz,  the closed string spacetime action reduces to Eqs. (\ref{eq: Odd with alpha x}-\ref{eq:coeff relation}). The derivations are completely parallel to those in Ref. \cite{Hohm:2019jgu}. Extra minus signs show up in the coefficients $c_k$ of the $\alpha'$ expansion.


\subsection*{Zeroth order  of $\alpha^{\prime}$}

\noindent We start with the tree-level closed string spacetime action without
$\alpha^{\prime}$ corrections

\begin{equation}
I_{0}\equiv\int d^{D}x\sqrt{-g}e^{-2\phi}\left[R+4\left(\partial_{\mu}\phi\right)^{2}-\frac{1}{12}H_{ijk}H^{ijk}\right],\label{eq:Polyakov}
\end{equation}

\noindent where $g_{\mu\nu}$ is the string metric, $\phi$ is the
dilaton and $H_{ijk}=3\partial_{\left[i\right.}b_{\left.jk\right]}$
is the field strength of the antisymmetric Kalb-Ramond field $b_{ij}$.
The ansatz  we use is

\begin{equation}
ds^{2}=-a^{2}\left(x\right)dt^{2}+dx^{2}+a^{2}\left(x\right)\left(dy^{2}+dz^{2}+\ldots\right),\quad b_{x\mu}=0,\label{eq:our ansatz}
\end{equation}

\noindent or

\begin{equation}
g_{\mu\nu}=\left(\begin{array}{ccc}
-a^{2}\left(x\right) & 0 & 0\\
0 & 1 & 0\\
0 & 0 & a^{2}\left(x\right)\delta_{ab}
\end{array}\right),\qquad b_{\mu\nu}=\left(\begin{array}{ccc}
0 & 0 & b_{0b}\left(x\right)\\
0 & 0 & 0\\
b_{a0}\left(x\right) & 0 & b_{ab}\left(x\right)
\end{array}\right),\qquad\phi=\phi\left(x\right),
\label{eq:our ansatz matrix}
\end{equation}

\noindent where $a,b=2,3,\ldots$ Mimicking the metric of cosmological backgrounds, we choose $b_{x\mu}=0$  in our ansatz.
It turns out that this gauge is crucial to preserving the $O(d,d)$ symmetry. In order to obtain the reduced
action by using the ansatz (\ref{eq:our ansatz matrix}), we rotate between the time-like $t$ and the first space-like
$x$ directions and rewrite the metric and $b_{\mu\nu}$ as

\begin{equation}
g_{\mu\nu}=\left(\begin{array}{cc}
1 & 0\\
0 & G_{ij}\left(x\right)
\end{array}\right),\qquad b_{\mu\nu}=\left(\begin{array}{cc}
0 & 0\\
0 & B_{ij}\left(x\right)
\end{array}\right),\label{eq:set up 1}
\end{equation}

\noindent where

\begin{equation}
G_{ij}\left(x\right)\equiv\left(\begin{array}{cc}
-a^{2}\left(x\right) & 0\\
0 & a^{2}\left(x\right)\delta_{ab}
\end{array}\right),\qquad B_{ij}\left(x\right)\equiv\left(\begin{array}{cc}
0 & b_{0b}\left(x\right)\\
b_{a0}\left(x\right) & b_{ab}\left(x\right)
\end{array}\right).\label{eq:set up 2}
\end{equation}

\noindent So $g_{00}\equiv g_{xx}$, $g_{11}\equiv g_{tt}$ and $b_{00}\equiv b_{xx}$, $b_{11}\equiv b_{tt}$. Henceforth, we will use Eqs. (\ref{eq:set up 1}) and (\ref{eq:set up 2}) as the definitions for $g_{\mu\nu}$ and $b_{\mu\nu}$.

Since we only need to use $G_{ij}$ as a whole to discuss the $O(d,d)$ symmetry and not its components, the time-like minus sign  $G_{11}\left(x\right)=-a^{2}\left(x\right)$ does not show up until we calculate the reduced action.   Straightforwardly, the Ricci tensor is

\begin{eqnarray}
R_{x}^{\;\;x} & = & -\frac{1}{4}\mathrm{Tr}\left(G^{-1}G^{\prime}\right)^{2}-\frac{1}{2}\mathrm{Tr}\left(G^{-1}G^{\prime\prime}\right)-\frac{1}{2}\mathrm{Tr}\left(G^{\prime}{}^{-1}G^{\prime}\right),\nonumber \\
R_{t}^{\;\;t} & = & -\frac{1}{2}\left(G^{-1}G^{\prime\prime}\right)_{t}^{\;\;t}-\frac{1}{4}\left(G^{-1}G^{\prime}\right)_{t}^{\;\;t}\mathrm{Tr}\left(G^{-1}G^{\prime}\right)+\frac{1}{2}\left(G^{-1}G^{\prime}G^{-1}G^{\prime}\right)_{t}^{\;\;t},\nonumber \\
R_{a}^{\;\;b} & = & -\frac{1}{2}\left(G^{-1}G^{\prime\prime}\right)_{a}^{\;\;b}-\frac{1}{4}\left(G^{-1}G^{\prime}\right)_{a}^{\;\;b}\mathrm{Tr}\left(G^{-1}G^{\prime}\right)+\frac{1}{2}\left(G^{-1}G^{\prime}G^{-1}G^{\prime}\right)_{a}^{\;\;b},\label{eq:set up 3}
\end{eqnarray}

\noindent and

\begin{equation}
H_{\mu\nu\alpha}H^{\mu\nu\alpha}=3H_{0ij}H^{0ij}=3B_{ij}^{\prime}\left(G^{-1}B^{\prime}G^{-1}\right)^{ij}=-3\mathrm{Tr}\left(G^{-1}B^{\prime}\right)^{2}.\label{eq:set up 4}
\end{equation}

\noindent where we used the notation

\begin{equation}
G'^{-1} \equiv \frac{d}{dx}(G^{-1}),\quad\quad \mathrm{Tr}\left(G^{-1}G^{\prime}\right)=g^{\mu\nu}g_{\mu\nu}^{\prime}.
\end{equation}

\noindent We then introduce the $O\left(d,d\right)$-invariant dilaton $\Phi$, defined  as

\begin{eqnarray}
\Phi & \equiv & 2\phi-\frac{1}{2}\ln\left|\det g_{\mu\nu}\right|,\nonumber \\
\Phi^{\prime} & = & 2\phi^{\prime}-\frac{1}{2}\mathrm{Tr}\left(G^{-1}G^{\prime}\right).\label{eq:sign 1}
\end{eqnarray}

\noindent Therefore, the action (\ref{eq:Polyakov}) can be rewritten
as

\begin{eqnarray}
\bar I_{0} & = & \int dx\, e^{-\Phi}\left[{\Phi'}^2+\frac{1}{4}\mathrm{Tr}\left(G^{-1}G^{\prime}\right)^{2}-\frac{1}{2}\mathrm{Tr}\left(G^{\prime}{}^{-1}G^{\prime}\right)+\frac{1}{4}\mathrm{Tr}\left(G^{-1}B^{\prime}\right)\right.\nonumber \\
 &  & \left.-\mathrm{Tr}\left(G^{-1}G^{\prime\prime}\right)+{\Phi'}\mathrm{Tr}\left(G^{-1}G^{\prime}\right)\right],
\end{eqnarray}

\noindent where we replaced $\bar I_0$ with $I_0$ to indicate that we are working with the ansatz (\ref{eq:our ansatz}).  Using integration by parts,
\begin{equation}
\frac{d}{dx}\left[e^{-\Phi}\mathrm{Tr}\left(G^{-1}G^{\prime}\right)\right]=e^{-\Phi}\left[\mathrm{Tr}\left(G^{-1}G^{\prime\prime}\right)+\mathrm{Tr}\left(G^{\prime-1}G^{\prime}\right)-\Phi^{\prime}\mathrm{Tr}\left(G^{-1}G^{\prime}\right)\right],
\end{equation}
the action becomes

\begin{equation}
\bar{I}_{0}=\int dxe^{-\Phi}\left[\Phi^{\prime2}-\frac{1}{4}\mathrm{Tr}\left(G^{-1}G^{\prime}\right)^{2}+\frac{1}{4}\mathrm{Tr}\left(G^{-1}B^{\prime}\right)^{2}\right].
\end{equation}

\noindent  Moreover, we want to point out  the  sign differences
between our ansatz (\ref{eq:our ansatz}) and the time-dependent FLRW
metric:

\begin{equation}
\mathrm{sign}\left[R_{xx}\right]=\mathrm{sign}\left[\tilde{R}_{tt}\right],\qquad\mathrm{sign}\left[R_{tt}\right]=-\mathrm{sign}\left[\tilde{R}_{xx}\right],\qquad\mathrm{sign}\left[R_{ab}\right]=-\mathrm{sign}\left[\tilde{R}_{ab}\right].\label{eq:sign 2}
\end{equation}

\begin{equation}
\mathrm{sign}\left[H^{2}\right]=-\mathrm{sign}\left[\tilde{H}^{2}\right],\qquad\mathrm{sign}\left[H_{\mu\nu}^{2}\right]=-\mathrm{sign}\left[\tilde{H}_{\mu\nu}^{2}\right],\qquad\mathrm{sign}\left[\left(\partial\phi\right)^{2}\right]=-\mathrm{sign}\left[\left(\partial\tilde{\phi}\right)^{2}\right],\label{eq:sign 3}
\end{equation}

\noindent where $\tilde{A}$ represents the quantities calculated
in the time-dependent FLRW background. For the Ricci scalar,
we have

\begin{equation}
\mathrm{sign}\left[R\right]=-\mathrm{sign}\left[\tilde{R}\right].\label{eq:sign 4}
\end{equation}

\noindent Finally, the tree-level action (\ref{eq:Polyakov}) becomes
\begin{equation}
\bar{I}_{0}=\int dxe^{-\Phi}\left[\Phi^{\prime2}+\frac{1}{8}\mathrm{Tr}\left(M^{\prime}\eta\right)^{2}\right].\label{eq:0th action}
\end{equation}

\noindent where

\begin{equation}
M=\left(\begin{array}{cc}
G^{-1} & -G^{-1}B\\
BG^{-1} & G-BG^{-1}B
\end{array}\right),\qquad\eta=\left(\begin{array}{cc}
0 & I\\
I & 0
\end{array}\right).
\end{equation}

\noindent Since $\mathrm{Tr}\left(M^{\prime}\eta\right)^{2}=\mathrm{Tr}\left(M^{\prime}M^{\prime}{}^{-1}\right)$, it is easy to see that the tree-level action (\ref{eq:0th action}) is invariant under  $O\left(d,d,R\right)$
transformations,

\begin{equation}
\Phi\rightarrow\Phi,\quad M\rightarrow\tilde{M}=\Omega^{T}M\Omega,
\end{equation}

\noindent where $\Omega$ satisfies $\Omega^{T}\eta\Omega=\eta$.  Considering the gravitational background with $B_{ij}=0$ and a global
transformation by  $\eta\in O\left(d,d,R\right)$, we have

\begin{equation}
M=\left(\begin{array}{cc}
G^{-1} & 0\\
0 & G
\end{array}\right)\rightarrow\tilde{M}=\eta M\eta=\left(\begin{array}{cc}
G & 0\\
0 & G^{-1}
\end{array}\right).
\end{equation}

\noindent This is the space-dependent duality   corresponding to the scale-factor duality in the time-dependent FLRW background.

\subsection*{First-order correction of $\alpha^{\prime}$}

We now demonstrate that  the closed string spacetime action with the first-order $\alpha^{\prime}$ correction also possesses the standard $O(d,d)$ symmetry for our ansatz  (\ref{eq:our ansatz}). The action with the first-order $\alpha^{\prime}$ correction and vanishing Kalb-Ramond field is

\begin{equation}
I=\int d^{D}x\sqrt{-g}e^{-2\phi}\left[R+4\left(\partial_{\mu}\phi\right)^{2}-\alpha^{\prime}\lambda_{0} R_{\mu\nu\sigma\rho}R^{\mu\nu\sigma\rho}+\mathcal{O}\left(\alpha^{\prime2}\right)\right].
\end{equation}

\noindent With some $\alpha'$-corrected field  redefinitions  as in \cite{Meissner:1996sa}, it can be expressed with first-order derivatives as

\begin{eqnarray}
I & &=  \int d^{D}x\sqrt{-g}e^{-2\phi}\left[R+4\left(\partial_{\mu}\phi\right)^{2}\right]\nonumber \\
 &  & -\alpha^{\prime}\lambda_{0}\int d^{D}x\sqrt{-g}e^{-2\phi}\left[-R_{GB}^{2}+16\left(R^{\mu\nu}-\frac{1}{2}g^{\mu\nu}R\right)\partial_{\mu}\phi\partial_{\nu}\phi-16\partial^{2}\phi \left(\partial\phi\right)^{2}+16\left(\partial\phi\right)^{4}\right] +\mathcal{O} \left(\alpha^{\prime2}\right),
\end{eqnarray}

\noindent where $R_{GB}^{2}$ is the Gauss-Bonnet term:

\begin{equation}
R_{GB}^{2}=R_{\mu\nu\sigma\rho}R^{\mu\nu\sigma\rho}-4R_{\mu\nu}R^{\mu\nu}+R^{2}.
\end{equation}

\noindent By using our ansatz (\ref{eq:our ansatz}), as one can check directly,  the action turns out to be

\begin{equation}
\bar{I}=-\int dxe^{-\Phi}\left\{ -\Phi^{\prime2}-\frac{1}{8}\mathrm{Tr}\mathcal{M}^{\prime2}+\alpha^{\prime}\lambda_{0}\left[\frac{1}{16}\mathrm{Tr}\mathcal{M}^{\prime4}-\frac{1}{64}\left(\mathrm{Tr}\mathcal{M}^{\prime2}\right)^{2}-\frac{1}{4}\Phi^{\prime2}\mathrm{Tr}\mathcal{M}^{\prime2}-\frac{1}{3}\Phi^{\prime4}\right]\right\}+\mathcal{O} \left(\alpha^{\prime2}\right),
\label{eq:1st order O(d,d)}
\end{equation}

\noindent with

\begin{equation}
\mathcal{M}\equiv M\eta=\left(\begin{array}{cc}
0 & G^{-1}\\
G & 0
\end{array}\right).
\end{equation}

\noindent Now, let us introduce the Kalb-Ramond field to the action
with the first-order $\alpha^{\prime}$ correction, which is
given by

\begin{eqnarray}
I & = & \int d^{D}x\sqrt{-g}e^{-2\phi}\left[R+4\left(\partial_{\mu}\phi\right)^{2}-\frac{1}{12}H^{2}\right]\nonumber \\
 &  & -\alpha^{\prime}\lambda_{0}\int d^{D}x\sqrt{-g} e^{-2\phi}\left[-R_{GB}^{2}+16\left(R^{\mu\nu}-\frac{1}{2}g^{\mu\nu}R\right)\partial_{\mu}\phi\partial_{\nu} \phi-16\partial^{2}\phi\left(\partial\phi\right)^{2}+16\left(\partial\phi\right)^{4}\right.,\nonumber \\
 &  & +\frac{1}{2}\left(R^{\mu\nu\sigma\rho}H_{\mu\nu\alpha} H_{\sigma\rho}^{\quad\alpha}-2R^{\mu\nu}H_{\mu\nu}^{2}+\frac{1}{3} RH^{2}\right)-2\left(D^{\mu}\partial^{\nu}\phi H_{\mu\nu}^{2}-\frac{1}{3}\partial^{2}\phi H^{2}\right)\nonumber \\
 &  & \left.-\frac{2}{3}H^{2}\left(\partial\phi\right)^{2} - \frac{1}{24}H_{\mu\nu\lambda}H_{\quad\rho\alpha}^{\nu} H^{\rho\sigma\lambda}H_{\sigma}^{\quad\mu\alpha}+\frac{1}{8}H_{\mu\nu}^{2}H^{2\mu\nu}-\frac{1}{144}\left(H^{2}\right)^{2}\right]+\mathcal{O} \left(\alpha^{\prime2}\right).
\end{eqnarray}

\noindent Using Eqs. (\ref{eq:set up 1}), (\ref{eq:set up 2}),
(\ref{eq:set up 3}),  (\ref{eq:set up 4}) and (\ref{eq:sign 1}), the action above can be expressed in the $O\left(d,d\right)$-invariant form (\ref{eq:1st order O(d,d)}), but with

\begin{equation}
\mathcal{M}=\left(\begin{array}{cc}
BG^{-1} & G-BG^{-1}B\\
G^{-1} & -G^{-1}B
\end{array}\right),
\end{equation}
as one can verify by applying the EOM of $\mathcal{M}$ and $\phi$.

\noindent The action (\ref{eq:1st order O(d,d)}) can be further simplified by using the tree-level EOM of $\Phi$  from Eq. (\ref{eq:0th action}), which is

\begin{equation}
\Phi^{\prime2}+\frac{1}{8}\mathrm{Tr}\left(M^{\prime}\eta\right)^{2}=0\rightarrow\Phi^{\prime2}=-\frac{1}{8}\mathrm{Tr}\mathcal{M}^{\prime2}.\label{eq:redefine}
\end{equation}

\noindent Then the action is reduced to

\begin{equation}
\bar{I}=-\int dxe^{-\Phi}\left\{ -\Phi^{\prime2}-\frac{1}{8}\mathrm{Tr}\mathcal{M}^{\prime2}+\alpha^{\prime}\lambda_{0}\left[\frac{1}{16}\mathrm{Tr}\mathcal{M}^{\prime4}+\frac{1}{96}\left(\mathrm{Tr}\mathcal{M}^{\prime2}\right)^{2}\right]\right\}.
\end{equation}

\noindent This action manifests the invariance under the $O\left(d,d,R\right)$
transformation
\begin{equation}
\Phi\rightarrow\Phi,\quad M\rightarrow\tilde{M}=\Omega^{T}M\Omega.
\end{equation}

\subsection*{Higher-order corrections of $\alpha^{\prime}$}
For our ansatz, we have shown that the zeroth order and the first
order in $\alpha^{\prime}$  can be rewritten in a standard $O\left(d,d\right)$-invariant form. Following the same logic as Refs. \cite{Hohm:2019ccp,Hohm:2019jgu}, it is reasonable to assume that this is also true for all orders in $\alpha'$.

Following the derivations in Ref. \cite{Hohm:2019jgu}, we now show that for our ansatz the action can be put into the reduced of (\ref{eq: Odd with alpha x}). From the definition $\mathcal{M}$, it is easy to get

\begin{equation}
\mathrm{Tr}\mathcal{M}=\mathrm{Tr}\mathcal{M}^{\prime}=\mathrm{Tr}\mathcal{M}^{\prime\prime}=0.
\end{equation}

\noindent Moreover,  $\mathcal{M}\mathcal{M}=1$ leads to   $\mathcal{M}\mathcal{M}^{\prime}+\mathcal{M}^{\prime}\mathcal{M}=0$ and then

\begin{equation}
2\mathcal{M}^{\prime}\mathcal{M}^{\prime}+\mathcal{M}^{\prime\prime}\mathcal{M}+ \mathcal{M}\mathcal{M}^{\prime\prime}=0\quad{\rm and} \quad \mathcal{M}\mathcal{M}^{\prime2k+1}=-\mathcal{M}^{\prime2k+1}\mathcal{M}.
\end{equation}

\noindent  Multiplying  by $\left(\mathcal{M}^{\prime}\right)^{2k+1}$   and taking traces,  one finds

\begin{equation}
\mathrm{Tr}\left(\mathcal{M}^{\prime2k+1}\right)=0,\qquad k=0,1,\ldots.
\end{equation}

\noindent Second, by using the equations of motion of the action
(\ref{eq:0th action}), higher space-dependent derivatives of $\mathcal{M}$
can be written in the terms of $\mathcal{M}$ and $\mathcal{M}^{\prime}$.
Therefore, higher-order corrections of $\alpha^{\prime}$ can be
built from $\mathcal{M}$ and $\mathcal{M}^{\prime}$. Third, if the
terms of the higher-order $\alpha^{\prime}$ corrections take the form
$\mathrm{Tr}\left(\mathcal{M}^{m}\mathcal{M}^{\prime k}\right)$,
by using $\mathcal{M}\mathcal{M}=1$ we  get

\begin{equation}
\mathrm{Tr}\left(\mathcal{M}\mathcal{M}^{\prime k}\right)=-\mathrm{Tr}\left(\mathcal{M}\mathcal{M}^{\prime k}\right)=0.
\end{equation}

\noindent Finally, due to Eq. (\ref{eq:redefine}),
the dilation could be replaced by $\mathrm{Tr}\mathcal{M}^{\prime2}$.
In summary, the higher-order corrections of $\alpha^{\prime}$ are
constructed using $\mathcal{M}^{\prime2}$. For example,

\begin{eqnarray}
\mathcal{O}\left(\alpha^{\prime}\right): &  & a_{1}\mathrm{Tr}\mathcal{M}^{\prime4}+a_{2}\left(\mathrm{Tr}\mathcal{M}^{\prime2}\right)^{2},\nonumber \\
\mathcal{O}\left(\alpha^{\prime2}\right): &  & b_{1}\mathrm{Tr}\mathcal{M}^{\prime6}+b_{2}\mathrm{Tr}\mathcal{M}^{\prime4}\mathrm{Tr}\mathcal{M}^{\prime2}+b_{3}\left(\mathrm{Tr}\mathcal{M}^{\prime2}\right)^{3},\nonumber \\
 & \vdots\nonumber \\
\mathcal{O}\left(\alpha^{\prime k-1}\right): &  & d_{1}\mathrm{Tr}\mathcal{M}^{\prime2k}+d_{2}\mathrm{Tr}\mathcal{M}^{\prime2k-2}\mathrm{Tr}\mathcal{M}^{\prime2}+d_{3}\mathrm{Tr}\mathcal{M}^{\prime2k-4}\left(\mathrm{Tr}\mathcal{M}^{\prime2}\right)^{2}+\ldots.
\end{eqnarray}

\noindent Furthermore, considering the action at zeroth order Eq.(\ref{eq:0th action}),
the variation for $g_{xx}$ gives

\begin{equation}
\delta\bar{I}_{0}=\int dxe^{-\Phi}\left[\Phi^{\prime2}+\frac{1}{8}\mathrm{Tr}\left(\mathcal{M}^{\prime}\right)^{2}\right]\delta g_{xx}.\label{eq:delta 1}
\end{equation}

\noindent This variation can be generalized to the higher orders with
$X_{2k}\left(\mathcal{M}^{\prime}\right)=\mathrm{Tr}\left[\left(\mathcal{M}^{\prime}\right)^{2k_{1}}\right]\cdots\mathrm{Tr}\left[\left(\mathcal{M}^{\prime}\right)^{2k_{n}}\right]$,
$k=k_{1}+k_{n}$:

\begin{equation}
\delta\bar{I}_{k}=\frac{\beta k}{2\left(4k-1\right)}\int dxe^{-\Phi}X_{2k}\left(\mathcal{M}^{\prime}\right)\mathrm{Tr}\left(\mathcal{M}^{\prime}\right)^{2},\label{eq:delta 2}
\end{equation}

\noindent where

\begin{equation}
\delta g_{xx}=\beta\alpha^{\prime k}X_{2k}\left(\mathcal{M}^{\prime}\right).\label{eq: rede gxx}
\end{equation}

\noindent If we substitute the redefinition (\ref{eq: rede gxx})
back into Eq. (\ref{eq:delta 1}) and set $\frac{\beta k}{2\left(4k-1\right)}=-1$,
we find that the terms with $\mathrm{Tr}\mathcal{M}^{\prime2}$ can
be eliminated when we sum Eqs. (\ref{eq:delta 1}) and (\ref{eq:delta 2}).
In other words, we can safely set  $\mathrm{Tr}\mathcal{M}^{\prime2}=0$ for $\alpha'$-corrected terms in the action
and obtain

\begin{eqnarray}
\mathcal{O}\left(\alpha^{\prime}\right): &  & a_{1}\mathrm{Tr}\mathcal{M}^{\prime4},\nonumber \\
\mathcal{O}\left(\alpha^{\prime2}\right): &  & b_{1}\mathrm{Tr}\mathcal{M}^{\prime6},\nonumber \\
 & \vdots\nonumber \\
\mathcal{O}\left(\alpha^{\prime k-1}\right): &  & d_{1}\mathrm{Tr}\mathcal{M}^{\prime2k}+d_{4}\mathrm{Tr}\mathcal{M}^{\prime2k-4}\mathrm{Tr}\mathcal{M}^{\prime4}\ldots.
\end{eqnarray}

\noindent The action with higher-order $\alpha^{\prime}$ corrections
then reduces to

\begin{equation}
\bar{I}\equiv-\int dxe^{-\Phi}\left(-\Phi^{\prime2}+{\sum_{k=1}^{\infty}}\left(\alpha^{\prime}\right)^{k-1}\bar{c}_{k}\mathrm{Tr}\left(\mathcal{M}^{\prime2k}\right)+\mathrm{multitraces}\right).
\end{equation}

\noindent After extracting the overall minus sign of the action above,
the even orders of $\alpha^{\prime}$ corrections acquire a
minus sign. Since $\bar{c}_{k}$ is the coefficient of $\mathrm{Tr}\mathcal{M}^{\prime2k}$,
which is not modified by   $\Phi^{\prime k}\simeq\frac{1}{8}\left(\frac{k-1}{k-3}\right)\Phi^{\prime k-2}\mathrm{Tr}\mathcal{M}^{\prime2}$,
the values of $\left|c_{k}\right|$ and $\left|\bar{c}_{k}\right|$
are the same. Moreover,
by using Eqs. (\ref{eq:sign 1}), (\ref{eq:sign 2}), and (\ref{eq:sign 3})
to all orders, we  find $\bar{c}_{1}=c_{1}=-\frac{1}{8}$, $\bar{c}_{2k+1}=c_{2k+1}$, and
$\bar{c}_{2k}=-c_{2k}$.
It is worth noting that the relationships between $\bar{c}_{k}$ and $c_{k}$ are not changed after including the contributions of the multitrace terms.

\end{document}